\documentclass{PoS}

\newcommand\ROBAST{\texttt{ROBAST}}

\newcommand\ROOT{\texttt{ROOT}}

\newcommand\libGeom{\texttt{libGeom}}

\newcommand\TGeoVolume{\texttt{TGeoVolume}}

\newcommand\TGeoManager{\texttt{TGeoManager}}
\newcommand\AOpticsManager{\texttt{AOpticsManager}}
\newcommand\AMirror{\texttt{AMirror}}
\newcommand\ALens{\texttt{ALens}}
\newcommand\AGeoAsphericDisk{\texttt{AGeoAsphericDisk}}
\newcommand\AGeoWinstonConePoly{\texttt{AGeoWinstonConePoly}}
\newcommand\AGeoBezierPgon{\texttt{AGeoBezierPgon}}

\newcommand\AObscuration{\texttt{AObscuration}}
\newcommand\AFocalSurface{\texttt{AFocalSurface}}

\newcommand\CPP{\texttt{C++}}

\newcommand\Bezier{B\'{e}zier}

\title{ROBAST: Development of a Non-Sequential Ray-Tracing Simulation Library and its Applications in the Cherenkov Telescope Array}

\ShortTitle{ROBAST and its Applications in CTA}

\author{\speaker{Akira Okumura}$^{a,b}$, Koji Noda$^c$, and Cameron Rulten$^d$, for the CTA Consortium\footnote{Full consortium author list at http://cta-observatory.org/}\\
  \llap{$^a$}Solar-Terrestrial Environment Laboratory, Nagoya University, Furo-cho, Chikusa-ku, Nagoya, Aichi 464-8601, Japan\\
  \llap{$^b$}Max-Planck-Institut f\"{u}r Kernphysik, P.O. Box 103980, D 69029 Heidelberg, Germany\\
  \llap{$^c$}Max-Planck-Institut f\"{u}r Physik, F\"{o}hringer Ring 6, D 80805 M\"{u}nchen, Germany\\
  \llap{$^d$}Department of Physics and Astronomy, University of Minnesota, 116 Church Street, Minneapolis, MN, 55455, U.S.A.\\
  E-mail: \email{oxon@mac.com} (A.~O.)}

\abstract{We have developed a non-sequential ray-tracing simulation library, \texttt{ROot-BAsed Simulator for ray Tracing} (\ROBAST), which is aimed for wide use in optical simulations of cosmic-ray (CR) and gamma-ray telescopes. The library is written in \CPP\ and fully utilizes the geometry library of the \ROOT\ analysis framework. Despite the importance of optics simulations in CR experiments, no open-source software for ray-tracing simulations that can be widely used existed. To reduce the unnecessary effort demanded when different research groups develop multiple ray-tracing simulators, we have successfully used \ROBAST\ for many years to perform optics simulations for the Cherenkov Telescope Array (CTA). Among the proposed telescope designs for CTA, \ROBAST\ is currently being used for three telescopes: a Schwarzschild--Couder telescope, one of the Schwarzschild--Couder small-sized telescopes, and a large-sized telescope (LST). \ROBAST\ is also used for the simulations and the development of hexagonal light concentrators that has been proposed for the LST focal plane. By fully utilizing the ROOT geometry library with additional \ROBAST\ classes, building complex optics geometries that are typically used in CR experiments and ground-based gamma-ray telescopes is possible. We introduce \ROBAST\ and show several successful applications for CTA.}

\FullConference{The 34th International Cosmic Ray Conference,\\
		30 July- 6 August, 2015\\
		The Hague, The Netherlands}

\begin{document}

\section{Introduction}

The optical systems in air-fluorescence and atmospheric Cherenkov telescopes (hereafter collectively called cosmic-ray (CR) telescopes) are a significant determinant of the telescope performance parameters, such as energy threshold, angular resolution, effective area, and field-of-view (FoV). Recent studies on optical systems have improved the potential ability of CR telescopes. For instance, novel optical systems that use Fresnel lenses, aspherical lenses, or aspherical mirrors have been proposed for JEM-EUSO, the Ashra experiment, and the Cherenkov Telescope Array (CTA), respectively \cite{Marchi:2010:The-JEM-EUSO-Mission-and-Its-Challenging-Optics-Sy,Sasaki:2002:Design-of-UHECR-telescope-with,Aita:2008:Ashra-Mauna-Loa-Observatory-an,Vassiliev:2007:Wide-field-aplanatic-two-mirro,Actis:2011:Design-concepts-for-the-Cherenkov-Telesc}, allowing observations of ultra-high energy CRs or very-high-energy (VHE) gamma rays with wider FOVs and higher angular resolutions than what is currently available.

The ray-tracing technique is widely used to simulate CR telescopes not only in the design study phase of optical systems but also in Monte Carlo simulations of air-shower events. However, the simulation of modern designs of CR telescopes has become harder as they have become more complex, due to the adoption of segmented mirrors, lenses, and complex telescope structures. In addition, the hexagonal light concentrators that are frequently used in front of the photodetectors are also simulated using the ray-tracing technique. To simulate complex CR telescopes and light concentrators, a so-called non-sequential ray-tracing technique is often employed. This technique can easily simulate multiple reflections on segmented mirrors or in a light concentrator without requiring the surface order in advance.

In previous decades, various research groups and projects developed and used different ray-tracing programs to simulate the optical systems of CR telescopes. However, the software requirements that surfaced in the simulations of CR telescopes were very similar. Thus, the unnecessary effort exerted in the development of multiple similar ray-tracing simulators by different groups can be reduced by developing a common, open-source simulator that can be widely used in CR research.

We have developed a \CPP\ library indented for various optics studies of CR telescopes. We introduce the functionality of our software and its practical applications in CTA.

\section{ROOT-Based Simulator for Ray Tracing}

\texttt{ROot-BAsed Simulator for ray Tracing} (\ROBAST) is a \CPP\ library based on the \ROOT\ analysis framework \cite{Brun:1997:ROOT----An-object-oriented-data-analysis} and its geometry library \cite{Brun:2003:The-ROOT-geometry-package}. We have developed \ROBAST\ to provide non-sequential ray-tracing functionality with several geometrical shapes that are frequently used in CR telescopes \cite{Okumura:2011:Development-of-Non-sequential-Ray-tracing-Software,Okumura:2015:ROBAST:-Development-of-a-ROOT-Based-Ray-Tracing-Li}. \ROBAST\ is currently being developed as an open-source project. The source code and some tutorials are available on the \ROBAST\ web page\footnote{http://robast.github.io/}. Refer to \cite{Okumura:2015:ROBAST:-Development-of-a-ROOT-Based-Ray-Tracing-Li} for the complete details of \ROBAST.

\subsection{Non-Sequential Ray-Tracing Simulation}

\ROBAST\ provides the functionality of non-sequential ray-tracing simulations, which is suited to optics simulations for segmented mirrors and light concentrators. The ray-tracing engine of \ROBAST\ is heavily dependent on the particle tracking engine of the \ROOT\ geometry library (\libGeom). \libGeom\ calculates the coordinates and the momentum vector of a particle moving through the detector  as well as the normal vector to the geometry surface. While \libGeom\ does not have a dedicated tool for ray tracing (i.e., the tracking, reflections, and refractions of optical photons), the above calculations can be appropriated to track optical photon trajectories by adding some new classes.

The non-sequential ray-tracing functionality of \ROBAST\ has been realized by implementing some basic classes derived from those of \ROOT. The \AOpticsManager\ class derived from \TGeoManager\footnote{\ROBAST\ and \ROOT\ class names start with ``A'' and ``T'', respectively. } performs majority of the non-sequential ray tracing. The \ALens, \AMirror, \AFocalSurface\, and \AObscuration\ classes derived from \TGeoVolume\ are used to model lenses, mirrors, focal surfaces, and telescope structures, which absorb photons, respectively.

\subsection{Geometry Building}

The 3D geometry of optical components can be modeled and built with the primitive shape classes provided by \libGeom. However, some modern CR telescope designs employ aspherical mirrors or lenses, or compound parabolic concentrators (Winston cones), which cannot be modeled with \libGeom\ classes, and thus, we have implemented the \AGeoAsphericDisk\ and \AGeoWinstonConePoly\ classes to model aspherical lenses and mirrors and Winston cones, respectively. A variant light concentrator design formed with a \Bezier\ curve \cite{Okumura:2012:Optimization-of-the-collection-efficiency-of-a-hex} can be also modeled using the \AGeoBezierPgon\ class.

\section{The Cherenkov Telescope Array and \ROBAST\ Applications}
CTA is a next-generation ground-based VHE gamma-ray observatory. It aims to detect more than 1000 gamma-ray sources with improved sensitivity (by a factor of ten) than that with the current generation of telescopes \cite{Actis:2011:Design-concepts-for-the-Cherenkov-Telesc,Acharya:2013:Introducing-the-CTA-concept}. It will comprise two arrays to cover both the southern and northern hemispheres, and several telescope designs have been proposed to achieve broad energy coverage from $20$~GeV to $300$~TeV.

Large-sized telescopes (LSTs), medium-sized telescopes (MSTs), and small-sized telescopes (SSTs) cover different energy bands with different optics designs. LSTs will observe the $20$--$200$~GeV band with a segmented parabolic system, MSTs the $100$~GeV--$10$~TeV band with a modified Davies--Cotton (DC) system \cite{Davies:1957:Design-of-the-Quartermaster-So}, and SSTs the $1$--$300$~TeV band with three candidate designs (two Schwarzschild--Couder (SC) \cite{Vassiliev:2007:Wide-field-aplanatic-two-mirro} designs and one DC design). In addition, another SC design (SC Telescope, SCT) has been proposed to cover the $200$~GeV--$10$~TeV band as an extension of the DC MSTs. Among these six telescopes designs, \ROBAST\ is currently being used for simulations of three telescopes, an LST, an SCT, and one of the SC-SSTs (Gamma Cherenkov Telescope; GCT), mainly for performance studies, tolerance analysis, and shadowing evaluations.

\begin{figure}
  \begin{minipage}[t][0.38\hsize][t]{0.33\hsize}
    (a)
    \begin{center}
      \includegraphics[height=.95\hsize]{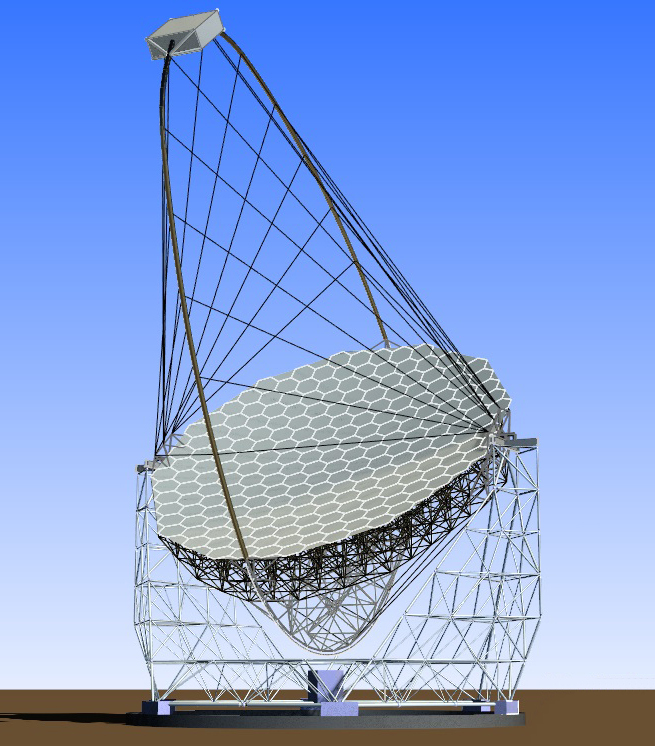}
    \end{center}
  \end{minipage}
  \begin{minipage}[t][0.38\hsize][t]{0.33\hsize}
    (b)
    \begin{center}
      \includegraphics[height=.95\hsize]{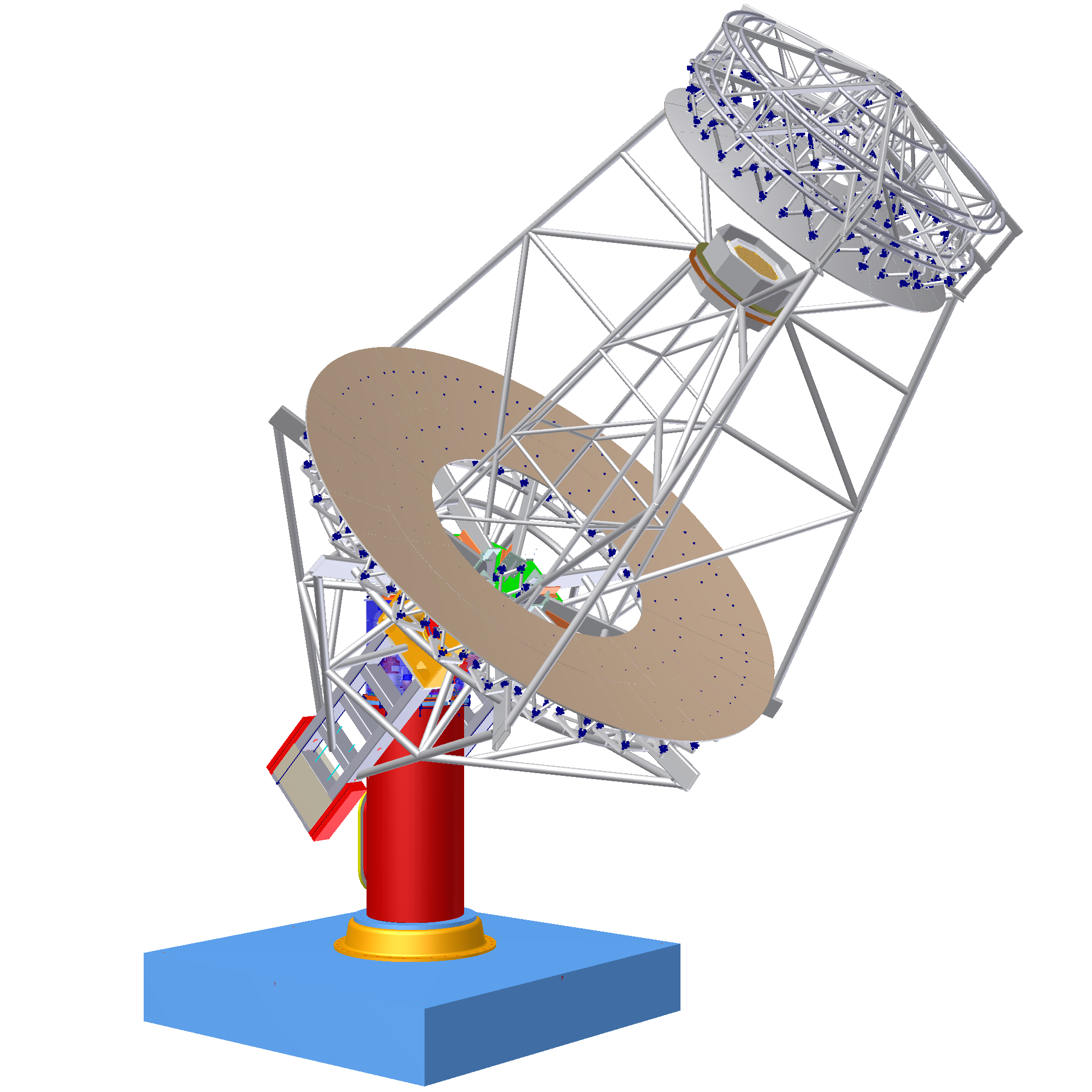}
    \end{center}
  \end{minipage}
  \begin{minipage}[t][0.38\hsize][t]{0.33\hsize}
    (c)\\ \\
    \begin{center}
      \includegraphics[width=\hsize]{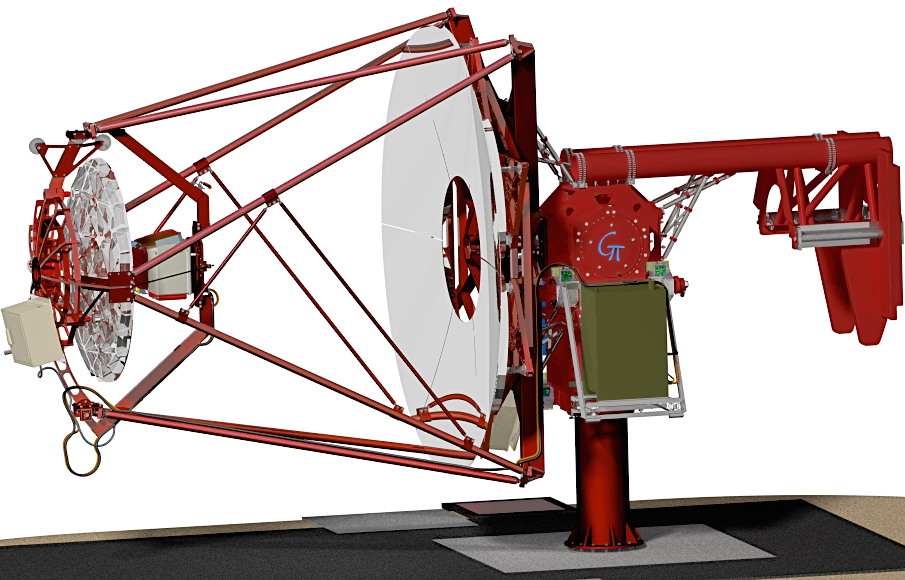}
    \end{center}
  \end{minipage}

  \begin{minipage}[t][0.38\hsize][t]{0.33\hsize}
    (d)
    \begin{center}
      \includegraphics[height=.98\hsize]{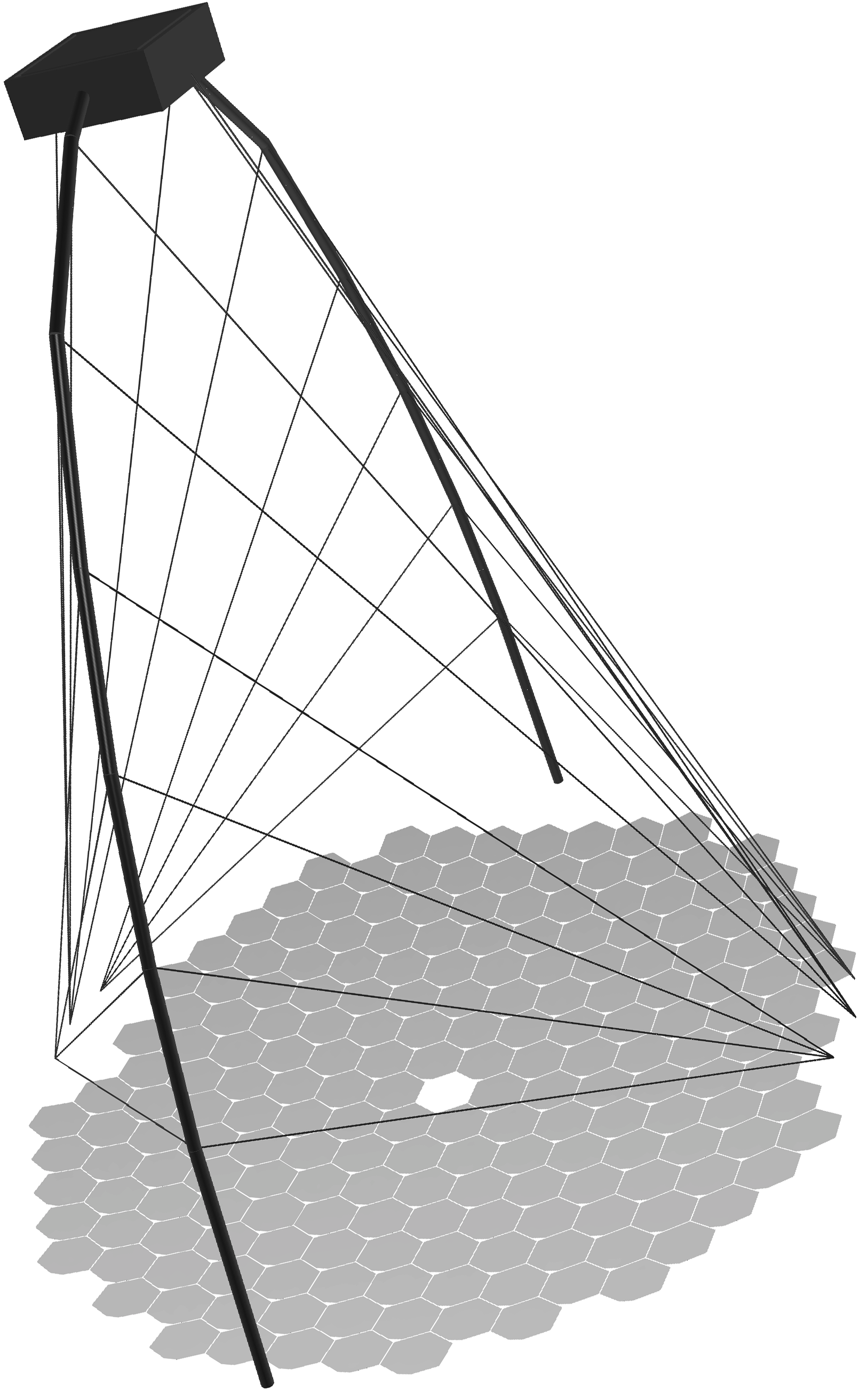}
    \end{center}
  \end{minipage}
  \begin{minipage}[t][0.38\hsize][t]{0.33\hsize}
    (e)
    \begin{center}
      \includegraphics[height=.98\hsize]{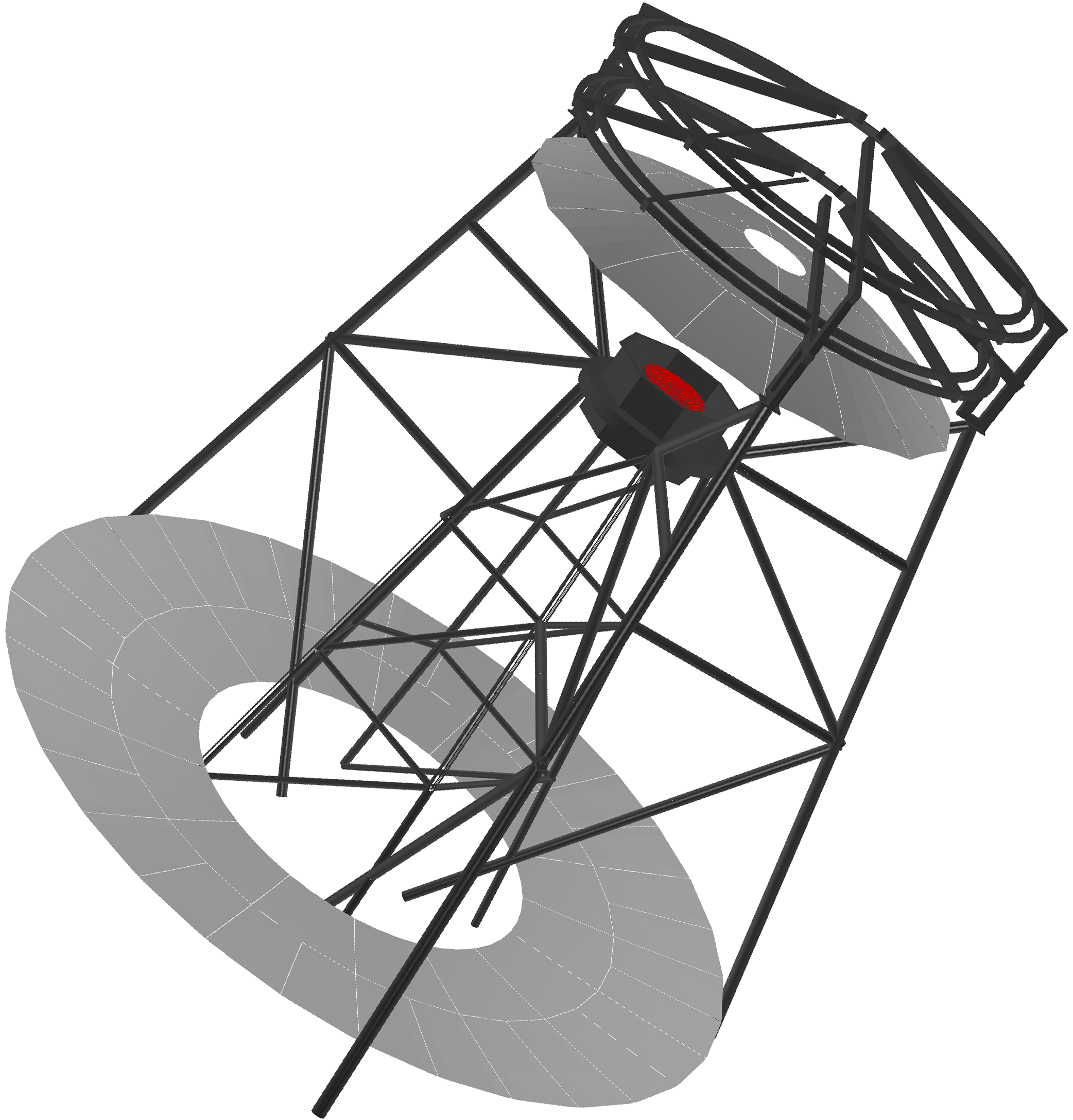}
    \end{center}
  \end{minipage}
  \begin{minipage}[t][0.38\hsize][t]{0.33\hsize}
    (f)
    \begin{center}
      \includegraphics[height=.8\hsize]{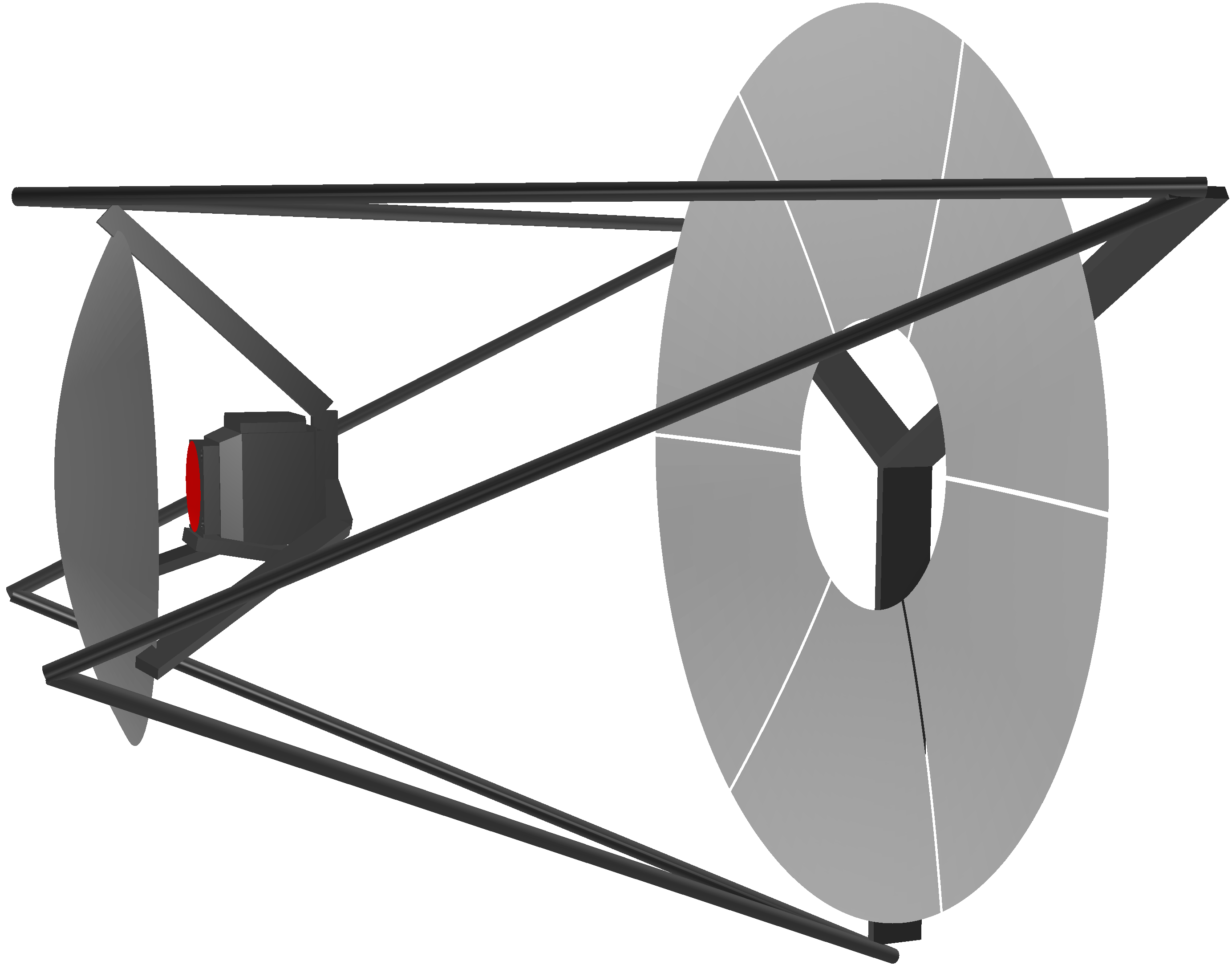}
    \end{center}
  \end{minipage}

  \begin{minipage}[t][0.32\hsize][t]{0.33\hsize}
    (g)
    \begin{center}
      \includegraphics[width=0.98\hsize]{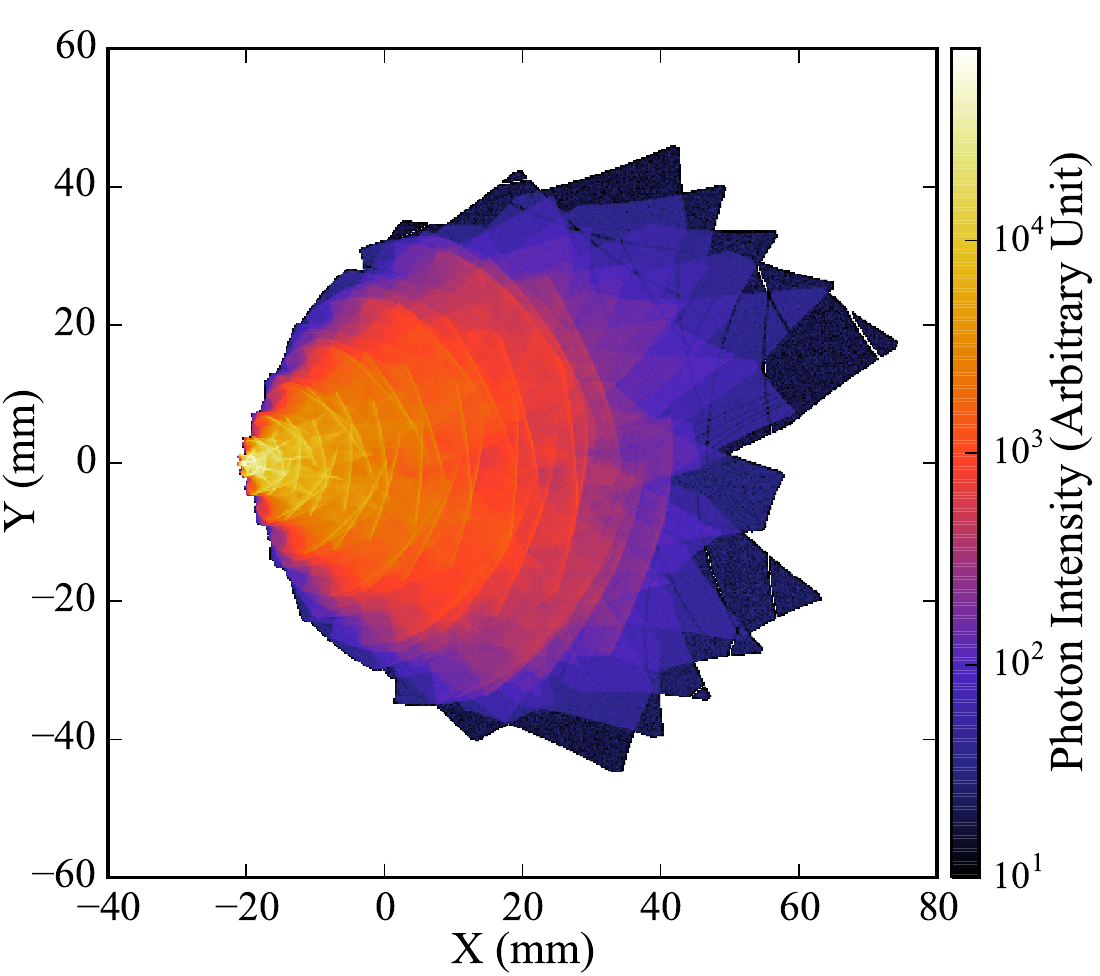}
    \end{center}
  \end{minipage}
  \begin{minipage}[t][0.32\hsize][t]{0.33\hsize}
    (h)
    \begin{center}
      \includegraphics[width=0.98\hsize]{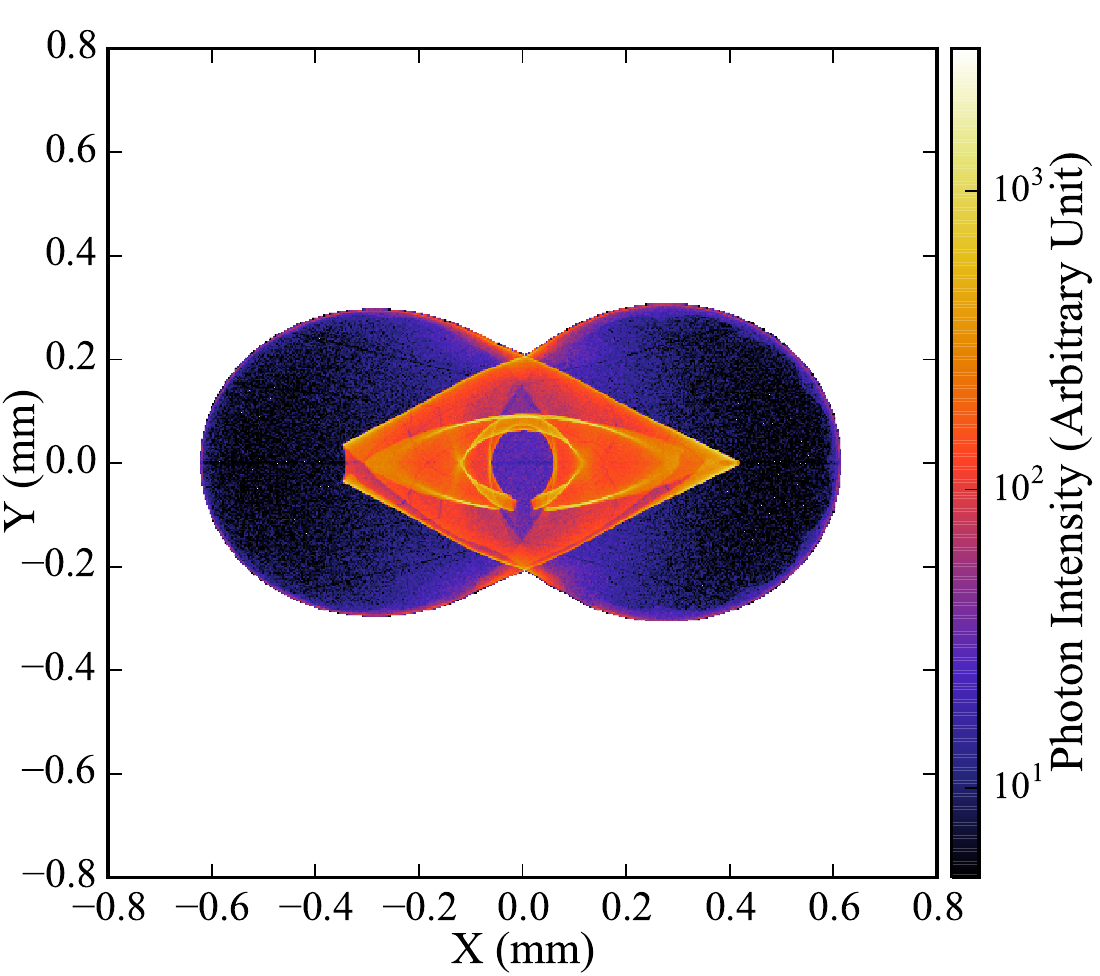}
    \end{center}
  \end{minipage}
  \begin{minipage}[t][0.32\hsize][t]{0.33\hsize}
    (i)
    \begin{center}
      \includegraphics[width=0.98\hsize]{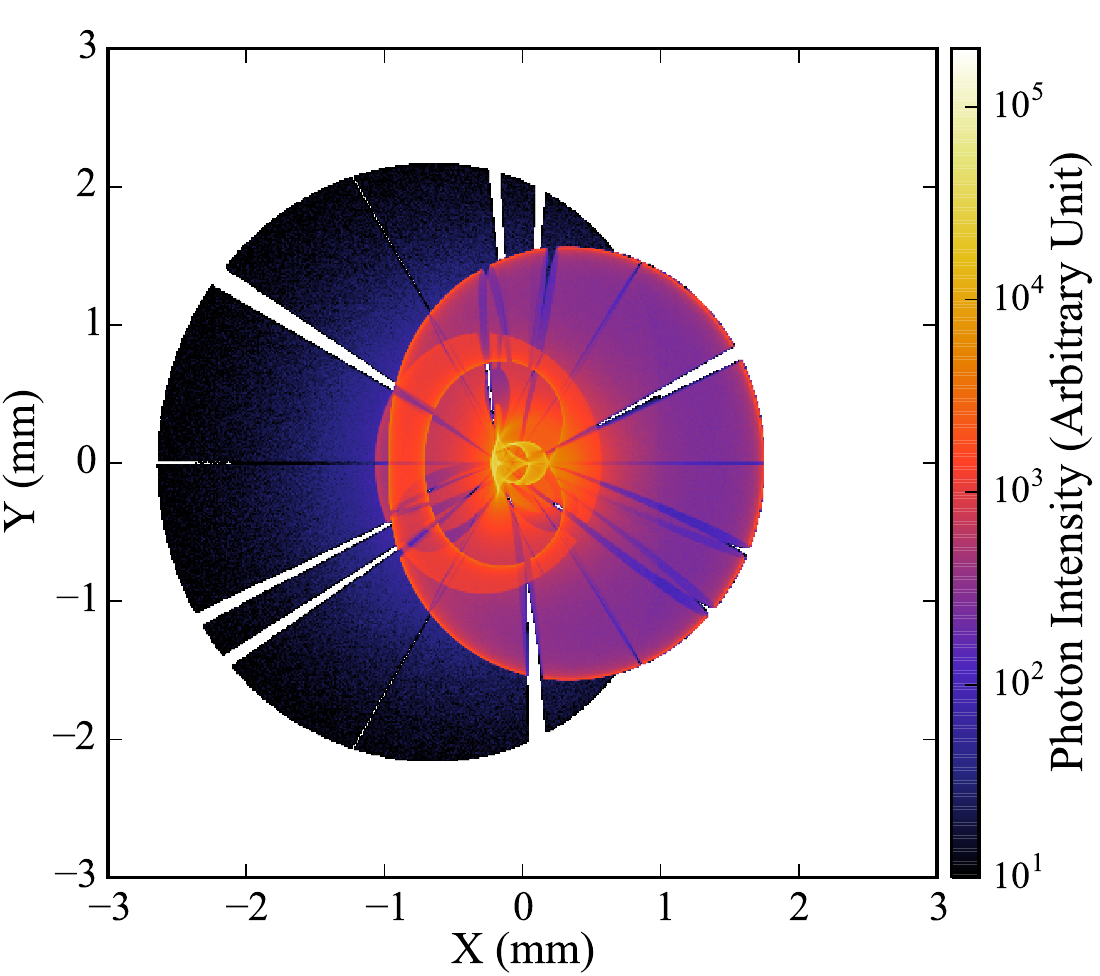}
    \end{center}
  \end{minipage}

  \caption{(a) 3D CAD model of an LST. (b) 3D CAD model of an SCT. (c) 3D CAD model of a GCT. (d)--(f) Equivalent \ROBAST\ geometries of (a)--(c). (g)--(i) Spot diagrams of the optical systems at a field angle of $1^\circ$. The spot diagrams shown here are for ideal optical systems and do not represent actual performance. Possible misalignments of the segmented mirrors or deviations in the mirror shapes have not been taken into account. (Image credit for figures (a)--(c): the CTA Consortium)}
    \label{fig:telescopes}
\end{figure}

Figures~\ref{fig:telescopes}(a)--(c) and (d)--(f) illustrate 3D CAD models of an LST, an SCT, and a GCT, and their equivalent \ROBAST\ models, respectively. The LST optical system comprising 198 segmented mirrors aligned on a parabola has been successfully modeled by \ROBAST. Its off-axis point spread function (PSF) at a field angle of $1^\circ$ (Figure~\ref{fig:telescopes}(g)) shows fine structure created by the segmented mirrors and the structure of the telescopes, demonstrating reflections and obscurations yielded by \AMirror\ and \AObscuration, respectively. Figures~\ref{fig:telescopes}(h) and (i) show off-axis PSFs of the SCT and GCT optical systems, respectively. These telescopes use SC optics designs comprising segmented aspherical primary and secondary mirrors that can be modeled with the \AGeoAsphericDisk\ class. Shadowing by the telescope masts and trusses as well as mirror segmentation can be seen in these spot diagrams.

\begin{figure}
  \begin{minipage}[t]{0.5\hsize}
    (a)
    \begin{center}
      \includegraphics[width=0.8\hsize]{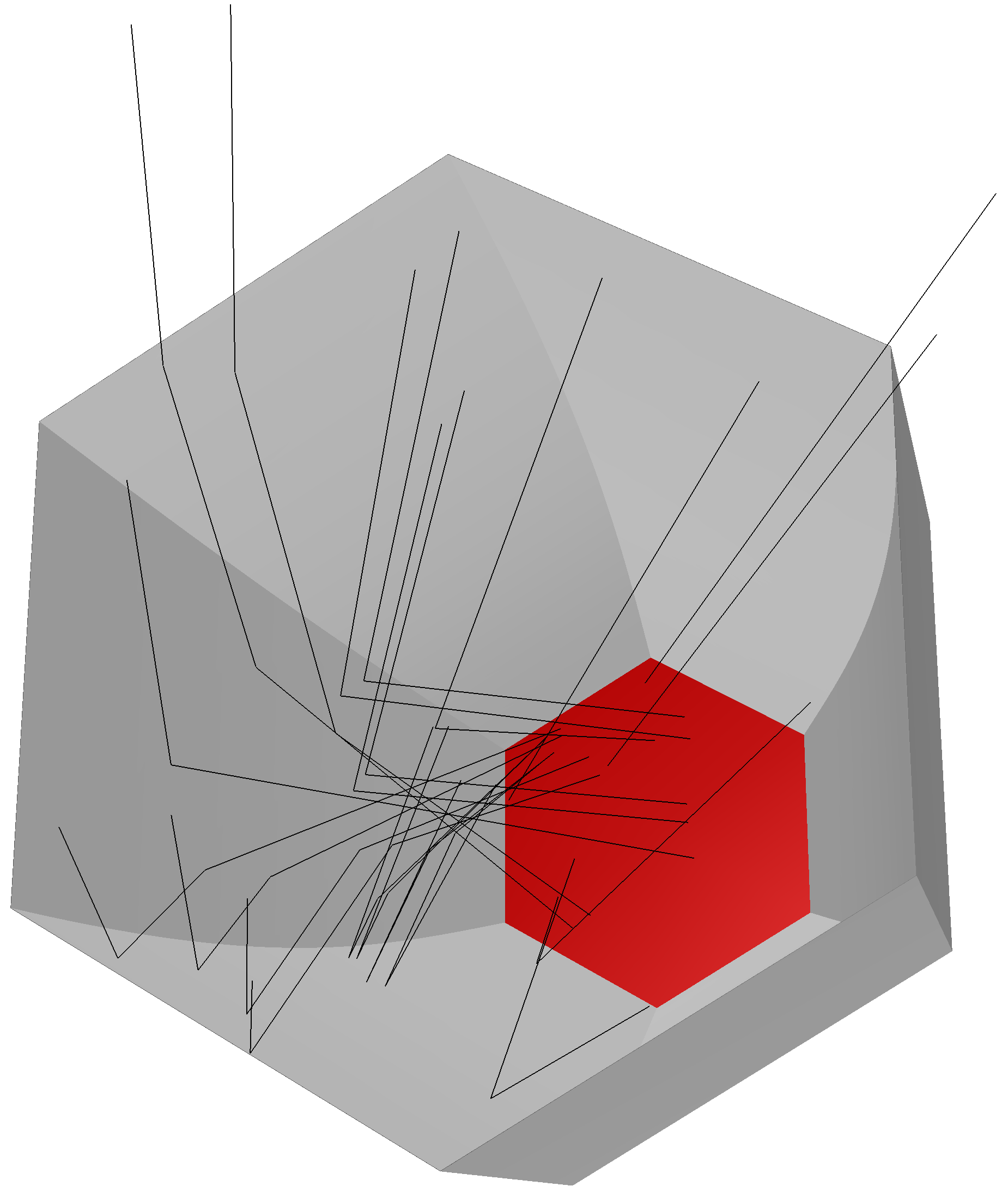}
    \end{center}
  \end{minipage}
  \begin{minipage}[t]{0.5\hsize}
    (b)
    \begin{center}
      \includegraphics[width=\hsize]{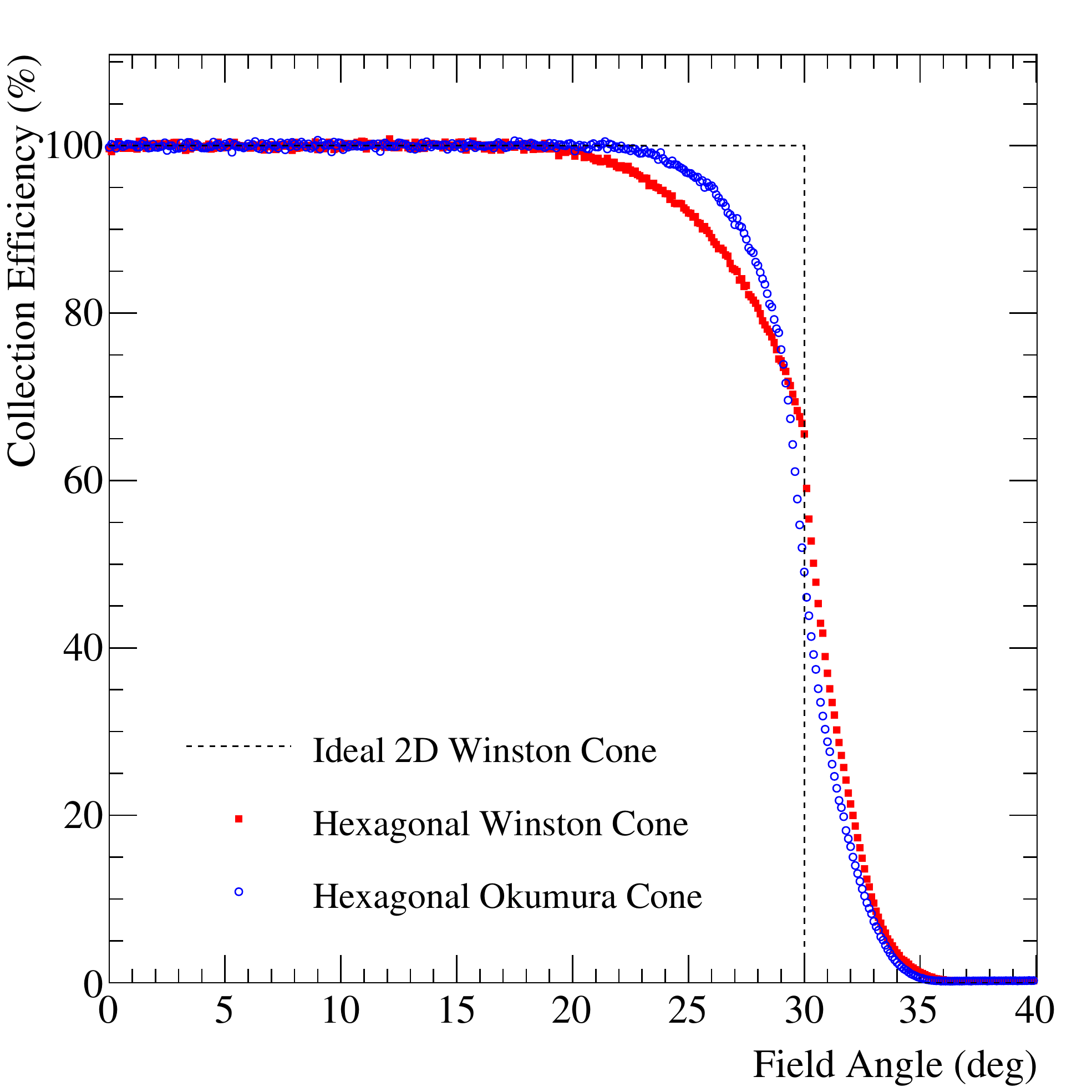}
    \end{center}
  \end{minipage}
  
  \caption{(a) \ROBAST\ model of a hexagonal Okumura cone. (b) Comparison of the collection efficiency of an ideal 2D Winston cone (black dashed), a hexagonal Winston cone (filled red squares), and a hexagonal Okumura cone (open blue circles).}
  \label{fig:Okumura}
\end{figure}

In addition to the simulation of telescopes, \ROBAST\ has been used for the design study of a hexagonal light concentrator for Cherenkov telescopes. In addition to conventional Winston cone designs, light concentrators with \Bezier\ curve surfaces (hereafter Okumura cones \cite{Okumura:2012:Optimization-of-the-collection-efficiency-of-a-hex}) can be modeled and simulated by \ROBAST. Figure~\ref{fig:Okumura}(a) illustrates a \ROBAST\ model of an Okumura cone and simulated photon tracks, revealing that the simulation of multiple reflections by the non-sequential ray-tracing method is important for light concentrators. Figure~\ref{fig:Okumura}(b) compares the collection efficiency of an ideal 2D Winston cone, a hexagonal Winston cone, and a hexagonal Okumura cone.

The LST camera comprising 1855 photomultiplier tubes will also use hexagonal light concentrators to increase the active area and reduce the stray light background \cite{Okumura:2015:Prototyping-of-Hexagonal-Light-Concentrators-for-t}. The design of the LST light concentrator is based on an Okumura cone, but its design has been optimized with position and incidence angle dependence of anode sensitivity being considered.

\section{Conclusion}
We have developed a non-sequential ray-tracing simulation library, \ROBAST,  as an open-source \CPP\ project. It is currently being used to study the optics of three CTA telescope designs and an LST light concentrator. Further details of \ROBAST\ will be reported in another upcoming paper \cite{Okumura:2015:ROBAST:-Development-of-a-ROOT-Based-Ray-Tracing-Li}.

\section*{Acknowledgments}
This study was supported by JSPS KAKENHI Grant Numbers 25610040 and 25707017. A.~O. was supported by a Grant-in-Aid for JSPS Fellows during his two-year visit to the University of Leicester, UK. We gratefully acknowledge support from the agencies and organizations listed under ``Funding Agencies'' at this website: http://www.cta-observatory.org/.

\bibliographystyle{ieeetr}
\bibliography{oxon}

%\begin{thebibliography}{99}
%\bibitem{...} 
%\end{thebibliography}

\end{document}